\documentclass[
    ,final            
  ]
  {aipproc}

\layoutstyle{6x9}

\begin{document}

\title{The SPS ion program and the first LHC data}

\classification{ 25.75.-q, 25.75.Ag, 25.75.Nq }
\keywords      {heavy ion collisions, onset of deconfinement, critical point}

\author{Marek Gazdzicki}{
  address={Goethe University Frankfurt, Germany and
           Jan Kochanowski University, Kielce, Poland}
}

\begin{abstract}
 For the first time in the CERN history two experimental programs
 devoted to study nucleus--nucleus collisions at high energies are
 performed in parallel. In the SPS ion program,  carried out by NA61/SHINE,
 interactions of light and medium size ions in the energy range 
 $\sqrt{s_{NN}}$~=~5--20~GeV are investigated. 
 The program aims to discover the critical point of strongly interacting 
 matter as well as establish properties of the onset of deconfinement.
 In 2010 ALICE, ATLAS and CMS at LHC recorded first data on Pb+Pb collisions 
 at the highest energy reached up to now, $\sqrt{s_{NN}}$~=~2760~GeV.
 This opens a new exciting area in the field of heavy ion collisions.
 The relation between the two programs is discussed
 in this presentation.
 Surprisingly, the first LHC results strongly support the NA49 discovery
 of the onset of deconfinement and thus further experimental study of
 nucleus-nucleus collisions at the CERN SPS.

\end{abstract}

\maketitle


\section{Introduction}

Experimental studies of nucleus-nucleus (A+A) collisions started
at the Super Proton Synchrotron (SPS) of the European Organization
for Nuclear Research (CERN) in the mid 1980s. 
They were motivated by the possibility to discover a new state of matter,
the quark-gluon plasma (QGP).
Firstly, beams of
oxygen and sulfur at 60$A$ and 200$A$~GeV/c were available.
Then, starting from the mid 1990s the lead beam at 158$A$~GeV/c
was used. These pioneering studies suggested that matter
of unusual properties is created at the early stage of A+A
collisions at the top SPS energy~\cite{Heinz}. Unambiguous evidence of
the QGP was, however, missing. This should be attributed
to the difficulty of obtaining unique predictions of the
QGP signals from the theory of strong interactions, the QCD.

For this reason, at the end of 1990, the NA49 Collaboration at
the CERN SPS started systematic search for the signals of the
onset of QGP creation.
This search was motivated by a statistical model of the early
stage of A+A collisions~\cite{GaGo} predicting that the onset of deconfinement
should lead to rapid changes  of the collision energy dependence 
of bulk properties of produced hadrons, all appearing in a common
energy domain.  
Data on central Pb+Pb collisions at 20$A$, 30$A$, 40$A$, 80$A$ and
158$A$ were recorded and the predicted features were observed at
low SPS energies~\cite{evidence,review}.

The NA49 evidence for the onset of deconfinement
motivates the ion program of NA61/SHINE~\cite{proposal} 
at the CERN SPS, as well as the beam energy scan at BNL RHIC and
the construction of the NICA ion collider at JINR, Dubna.
The basic goals of this experimental effort are the study
of the properties of the onset of deconfinement and the search
for the critical point of strongly interacting matter~\cite{proposal}.
Both relay on the correctness of the NA49 results and their
interpretation. 

Up to recently, the evidence for the onset of deconfinement
was based on data of
a single experiment, NA49 at the CERN SPS.
Thus, an independent verification of the relevant NA49 
measurements is important.
Furthermore, it is very crucial to confirm the interpretation
of the NA49 results in terms of
the onset of deconfinement,
firstly, for confirming the discovery of
the onset of deconfinement and, secondly, to strengthen
arguments for the NA61/SHINE and other experimental programs
with high energy ion beams.

This year rich data from the RHIC beam energy scan program 
were released~\cite{kumar}. They agree with the NA49 measurements relevant
for the onset of deconfinement. Furthermore, the first
results on Pb+Pb collisions at the CERN LHC were presented~\cite{schukraft}.
The latter strongly confirm the interpretation of the NA49 results as
an observation of the onset of deconfinement.    
This contribution summarizes the status of the evidence for the onset
of deconfinement including the new LHC and RHIC results.

\section{Status of the evidence for the onset of deconfinement}

\begin{figure}[!htb]
\begin{minipage}[b]{0.95\linewidth}
\includegraphics[width=0.5\linewidth]{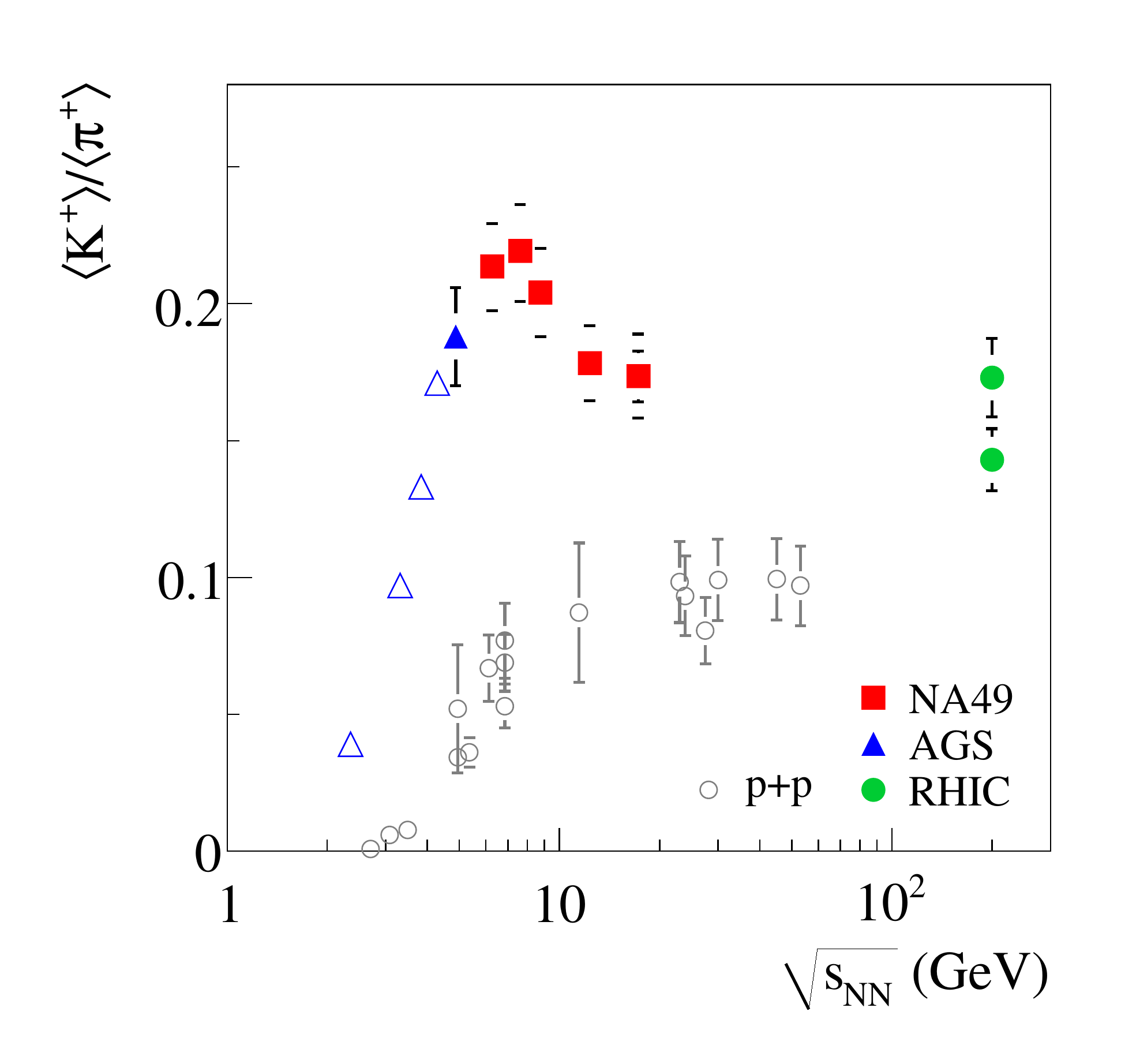}
\includegraphics[width=0.5\linewidth]{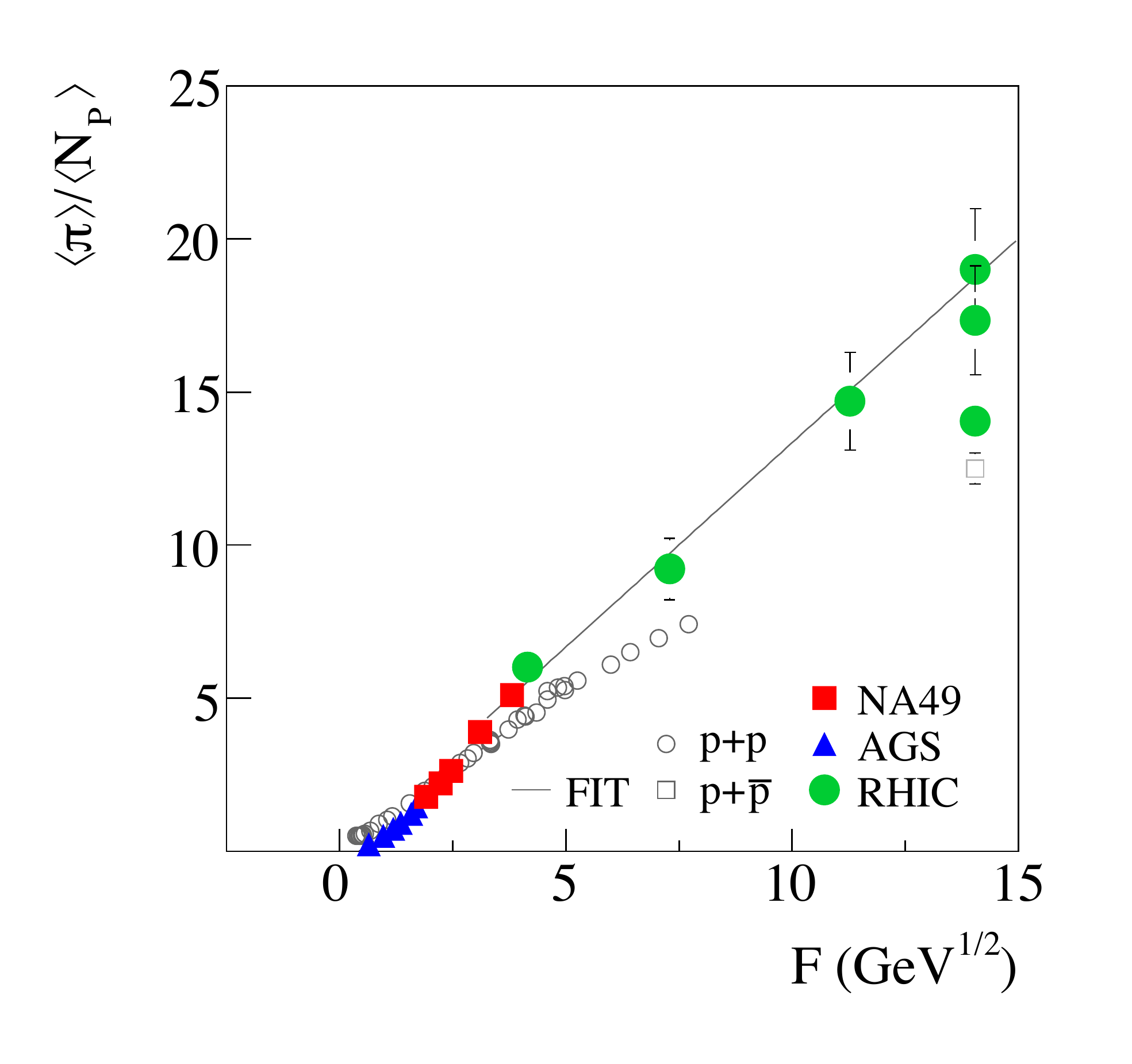}
\includegraphics[width=0.5\linewidth]{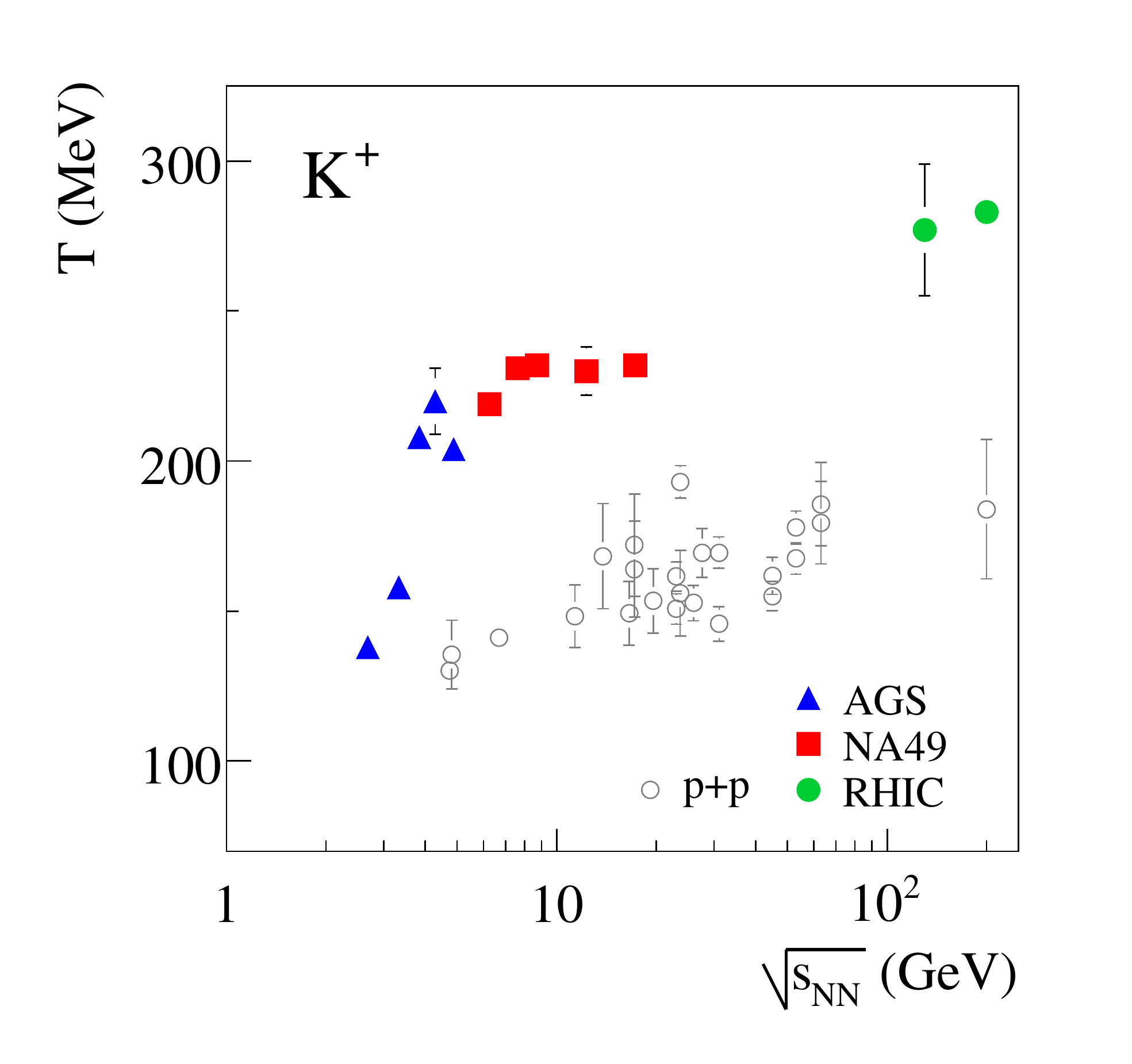}
{\hspace*{0.2 cm}
\includegraphics[width=0.5\linewidth]{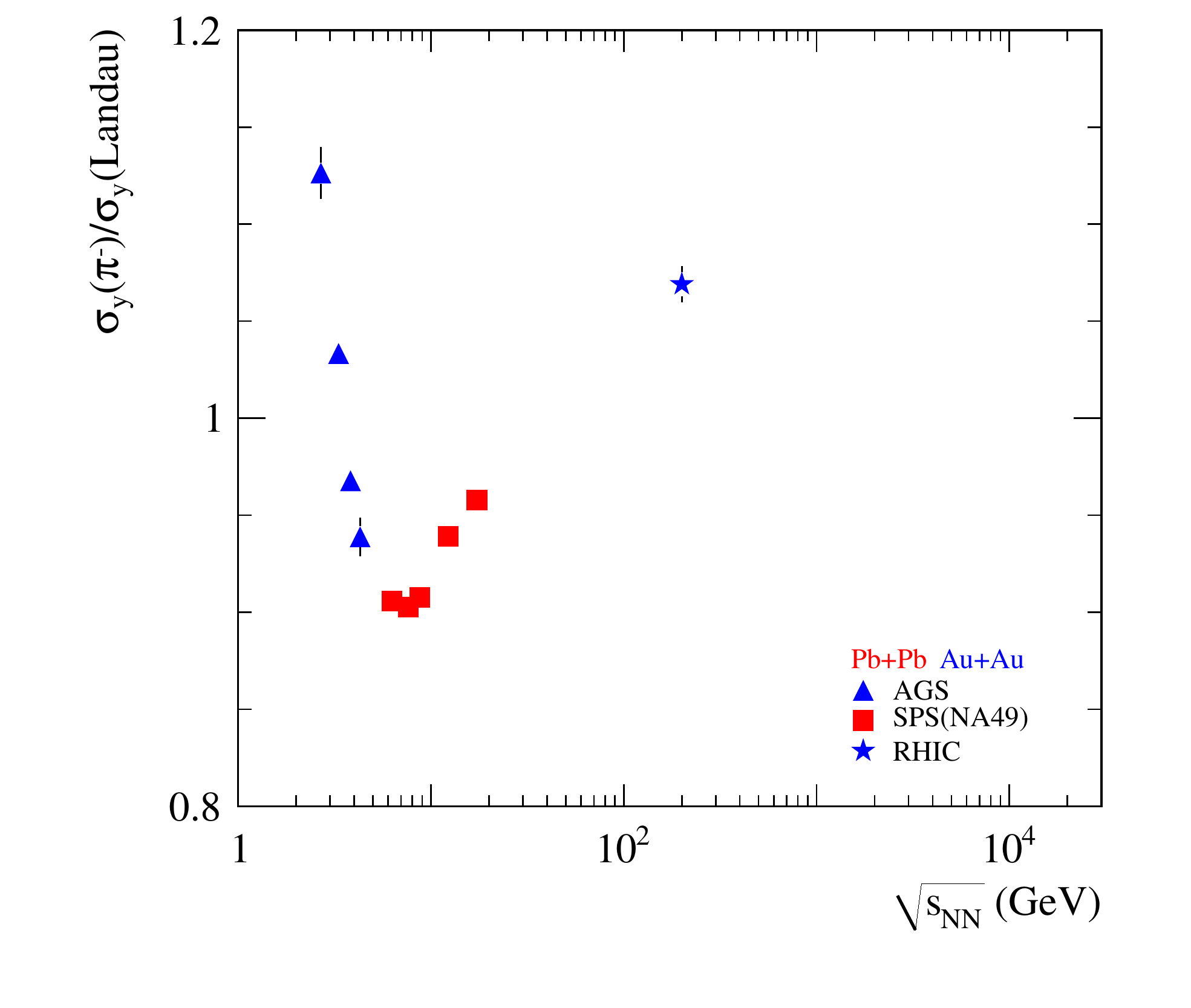}}
\end{minipage}
\caption{\label{heating_curves}
Heating curves of strongly interacting matter, status at the mid of 2010.
Hadron production properties (see Ref.~\cite{review} for details)
are plotted
as a function of collision energy ($\sqrt{s_{NN}}$ and
$F \approx \sqrt{\sqrt{s_{NN}}}$) for central Pb+Pb (Au+Au)
collisions and p+p interactions
(open circles): 
top-left -- the $\langle K^+ \rangle /\langle \pi^+\rangle$ ratio,
top-right -- the mean pion multiplicity per participant nucleon,
bottom-left -- the inverse slope parameter  of the
transverse mass spectra of $K^+$  mesons,
bottom-right: the width of the $\pi^-$ rapidity spectra relative
to predictions of the Landau ideal hydrodynamics.
The observed changes of the energy dependence for central Pb+Pb (Au+Au)
collisions are related to:
decrease of the mass of strangeness carriers and the ratio of
strange to non-strange degrees of freedom ({\it horn}: top-left plot),
increase of entropy production ({\it kink}: top-right plot),
weakening of transverse ({\it step}: bottom-left plot) and
longitudinal ({\it dale}: bottom-right plot)
expansion at the onset of deconfinement.
}
\end{figure}

The NA49 evidence for the onset of deconfinement~\cite{evidence,review} 
is based on the  observation
that numerous hadron production properties measured
in central Pb+Pb collisions change their energy dependence
in a common energy domain (starting from  
$\sqrt{s_{NN}} \approx$~7.6~GeV~($\approx$ 30$A$~GeV/c
beam momentum)) and
that these changes are consistent with the predictions
for the onset of deconfinement~\cite{GaGo}. The four representative
plots  with the structures
referred to as $horn$, $kink$, $step$ and $dale$~\cite{review}
are shown in Fig.~\ref{heating_curves}. They present the experimental
results available in the mid of 2010.

The relation between the $horn$, $kink$, $step$ and $dale$
structures and the onset of deconfinement is briefly
discussed below. More detailed explanation is given
in Ref.~\cite{review}, where  a comparison with quantitative
models is also presented.

{\it The horn.}
The most dramatic change of the energy dependence is seen for
the ratio of particle yields of kaons and pions, Fig.~\ref{heating_curves} (top-left).
The steep threshold rise of the ratio
characteristic for confined matter changes at high energy into a constant value
at the level expected for deconfined matter. In the transition
region (at low SPS energies) a sharp maximum is observed caused by
the higher strangeness to entropy production ratio in confined matter than
in deconfined matter.
This feature is not observed for proton--proton reactions as shown
by the open dots in Fig.~\ref{heating_curves} (top-left).

{\it The kink.}
The majority of all particles produced in high energy interactions
are pions. Thus, pions carry basic information on the entropy
created in the collisions. On the other hand, entropy production
should depend on the form of matter present at the early stage of
collisions. Deconfined matter is expected to lead to a final state
with higher entropy than that created by confined matter.
Consequently, the entropy  increase at the onset of deconfinement
is expected to lead  to a steeper increase of the collision energy dependence
of the pion yield per participating nucleon. This effect is observed
for central Pb+Pb collisions as shown in Fig.~\ref{heating_curves} (top-right).
When passing the low SPS energies the slope of the  
$\langle \pi \rangle / \langle N_P \rangle$ vs $F \approx \sqrt{\sqrt{s_{NN}}} $
dependence increases by a factor of about 1.3.
Within the statistical model of the early stage~\cite{GaGo} 
this corresponds to an
increase of the effective number of degrees of freedom by a factor of
about 3.

{\it The step.}
The experimental results on the energy dependence of the inverse
slope parameter, $T$, of $K^+$  and $K^-$ transverse mass spectra for
central Pb+Pb (Au+Au) collisions are shown in Fig.~\ref{heating_curves} (bottom-left).
The striking features of the data can be summarized and
interpreted~\cite{GoGaBu} as follows. The $T$ parameter
increases strongly with collision energy up to the low SPS energies,
where the creation of confined matter at the early stage of
the collisions takes place. 
In a pure phase increasing collision energy leads to
an increase of the early stage temperature and pressure.
Consequently the transverse momenta of produced hadrons, measured
by the inverse slope parameter, increase with collision energy.
This rise is followed by a region of approximately constant value
of the $T$ parameter in the SPS energy range,
where 
the transition between confined and deconfined matter with
the creation of mixed phases is located. The resulting softening of the
equation of state, EoS, `suppresses' the hydrodynamical transverse
expansion and leads to the observed plateau structure in the
energy dependence of the $T$ parameter \cite{GoGaBu}. At higher
energies (RHIC data), $T$ again increases with the collision
energy. The EoS at the early stage becomes again stiff and the
early stage pressure increases with collision energy, resulting in
a resumed increase of $T$.

{\it The dale.}
As discussed above, the weakening of the transverse expansion
is expected due to the onset of deconfinement because of the
softening of the EoS at the early stage. Clearly the latter 
should also weaken the longitudinal expansion. 
This expectation is checked in Fig.~\ref{heating_curves} (bottom-right),
where the width of the $\pi^-$ rapidity spectra in central Pb+Pb
collisions relative to predictions of the Landau ideal hydrodynamics is plotted as
a function of the collision energy.
In fact, the ratio has a clear minimum at low SPS energies.

\begin{figure}[!htb]
\begin{minipage}[b]{0.95\linewidth}
\includegraphics[width=0.5\linewidth]{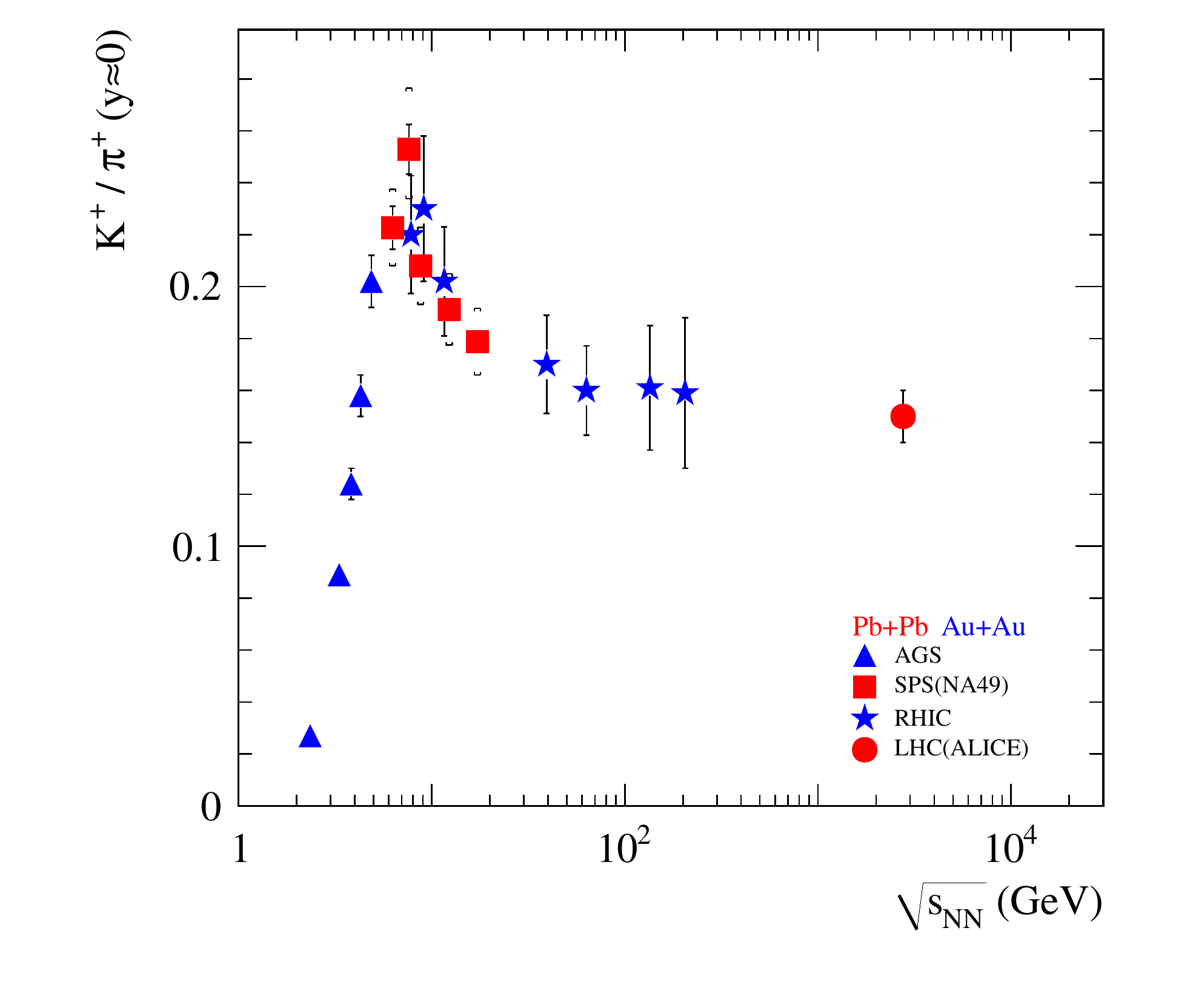}
\includegraphics[width=0.5\linewidth]{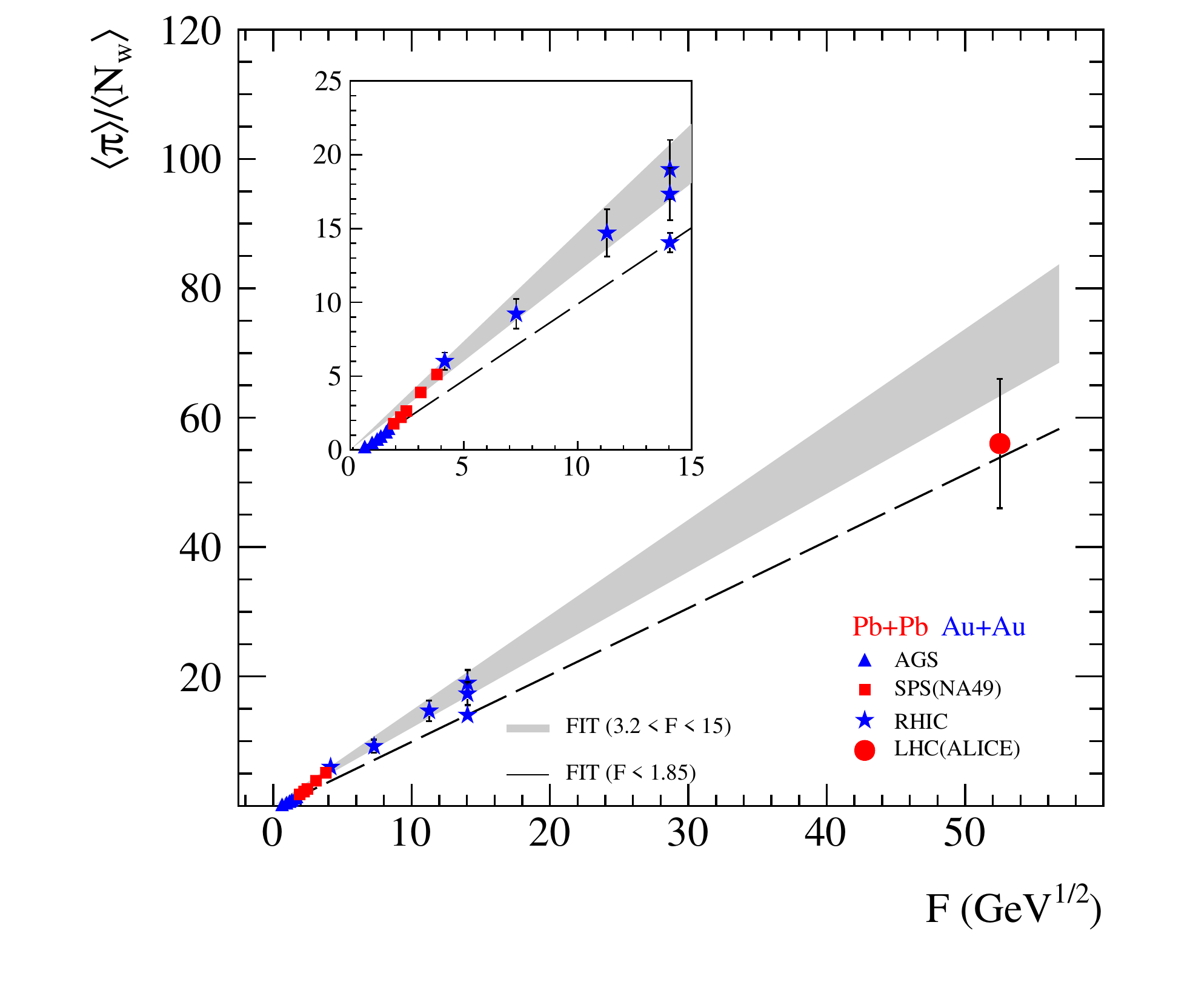}
\includegraphics[width=0.5\linewidth]{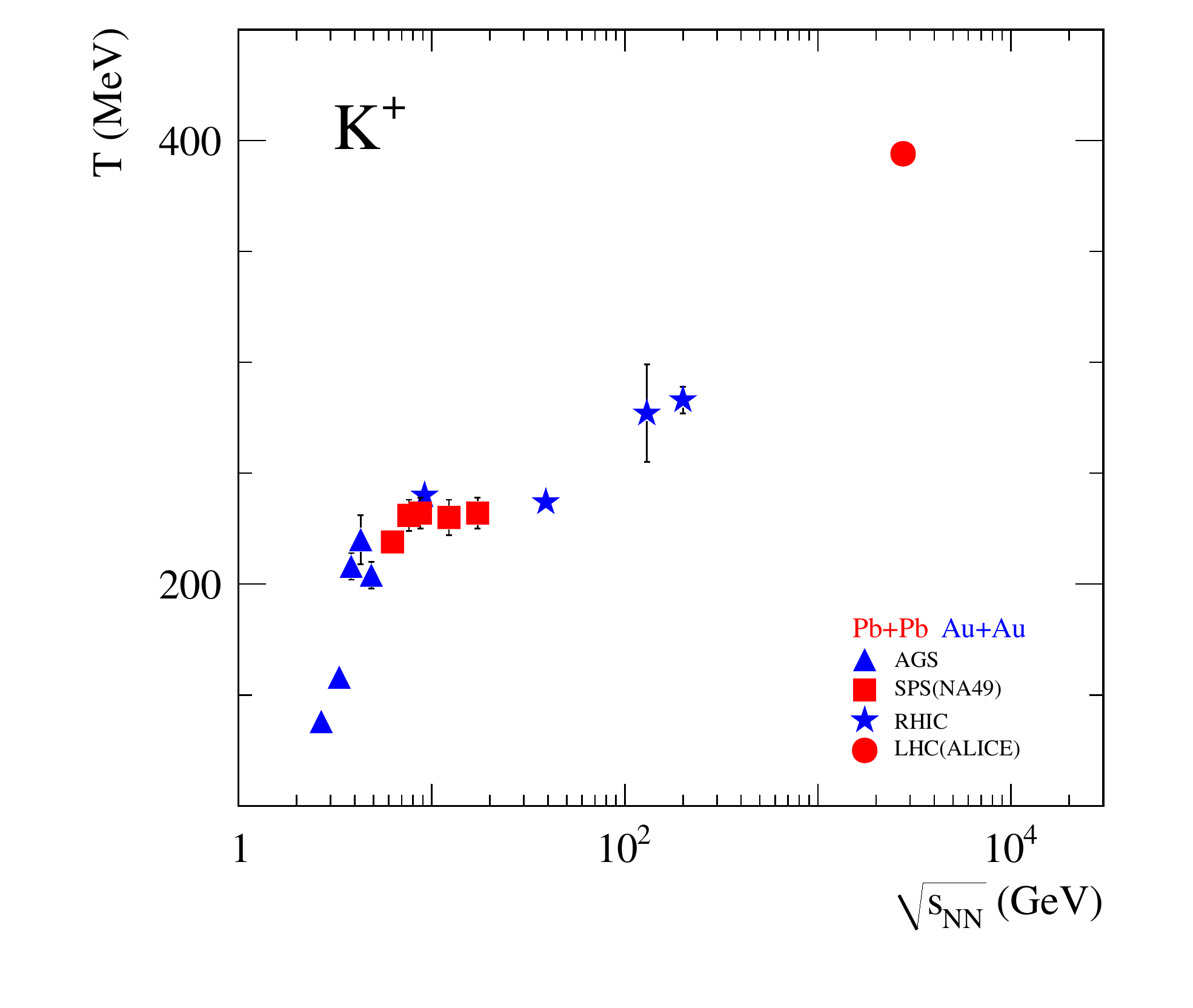}
{\hspace*{0.2 cm}
\includegraphics[width=0.5\linewidth]{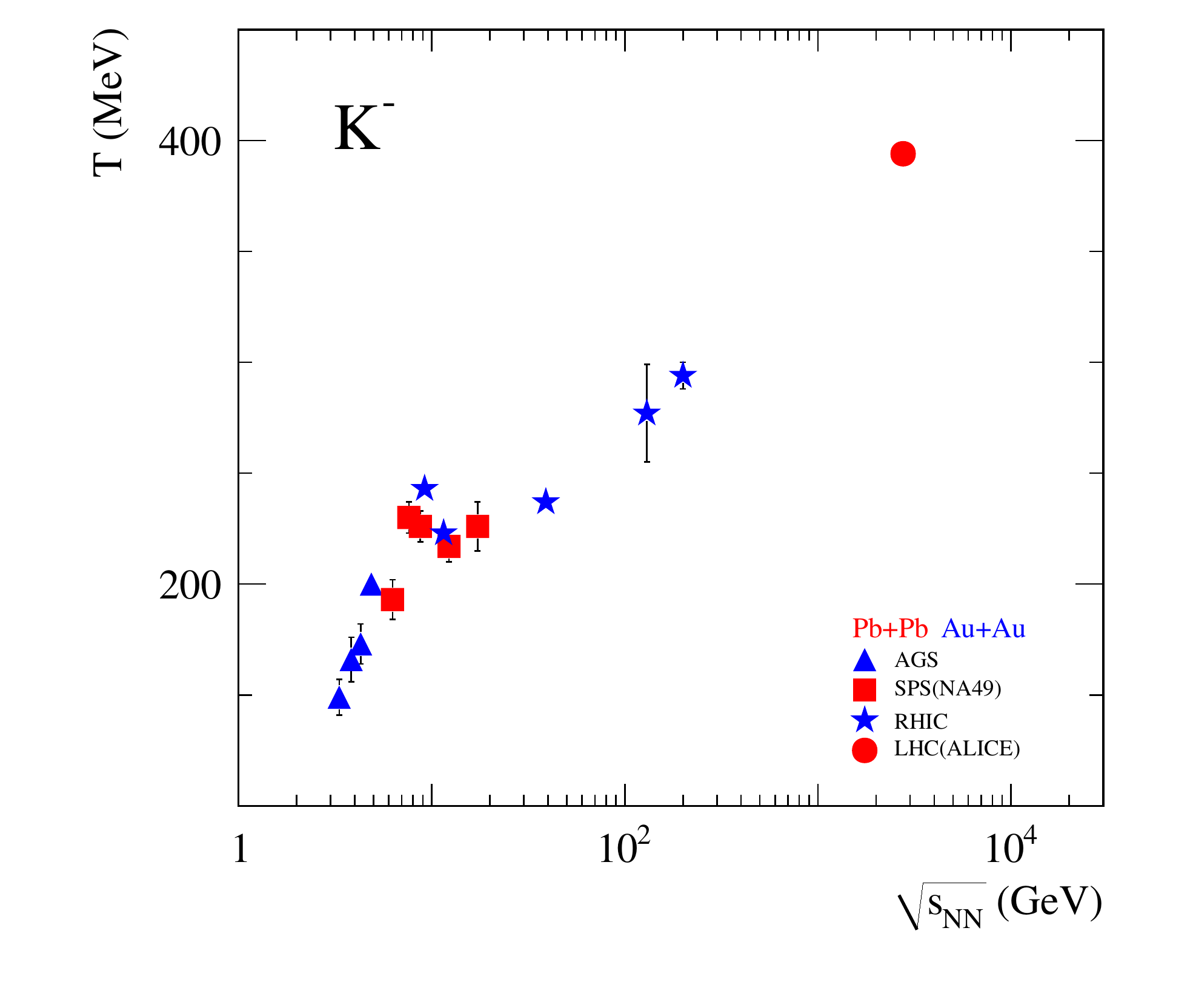}}
\end{minipage}
\caption{\label{evidence_2011}
Heating curves of strongly interacting matter: status  at the mid of 2011. 
For more details see see Ref.~\cite{review} and the caption of Fig.~\ref{heating_curves}. 
The new LHC and RHIC data are included in the 
{\it horn} (top-left), {\it kink} (top-right) and {\it step} (bottom)
plots. 
The $K^+/\pi^+$ ratio is measured by ALICE~\cite{schukraft} and
STAR~\cite{kumar} at mid-rapidity only and thus the {\it horn}
plot is shown here for the mid-rapidity data.
There are no new results for the {\it dale} plot.
}
\end{figure}

In 2011 new results on central Pb+Pb collisions at the LHC and
data on central Au+Au collisions from the RHIC beam energy scan
program were released. 
The updated plots~\cite{anar} are shown in Fig.~\ref{evidence_2011}. 
The RHIC results~\cite{kumar} agree with the NA49 measurements at the onset energies.

The LHC data~\cite{schukraft} 
demonstrate that the energy dependence of hadron production properties
shows rapid changes only at low SPS energies. A smooth evolution is observed
between the top SPS (17.2~GeV) and the current LHC (2.76~TeV) energies. 
This strongly supports 
the interpretation of the NA49 structures as due to the onset of deconfinement.
Above the onset energy only a smooth change of the
quark-gluon plasma properties with increasing collision energy is expected.
Consequently, in agreement with the first LHC data, one expects:
\begin{itemize}
\item
an approximate independence of the $K^+/\pi^+$ ratio of energy above the
the top SPS energy, Fig.~\ref{evidence_2011} (top-left),
\item
a linear increase of the pion yield per participant with $F$ with the
slope defined by the top SPS data, Fig.~\ref{evidence_2011} (top-right),
\item
a  monotonic increase of the kaon inverse slope parameter
with energy above the top SPS energy, Fig.~\ref{evidence_2011} (bottom).
\end{itemize}
The width of the $\pi^-$ rapidity spectra in central Pb+Pb
collisions relative to predictions of the Landau ideal hydrodynamics
should continuously increase from the top SPS to LHC energies.
The LHC data on rapidity spectra are needed to verify this expectation.

The confirmation of the relevant NA49
measurements and their interpretation in terms of
the onset of deconfinement by the new LHC and RHIC data
strengthen the arguments for the planned~\cite{proposal}
NA61/SHINE measurements
with secondary Be and primary Ar as well as Xe beams in the SPS 
beam momentum range (13$A$-158$A$~GeV/c).



\begin{theacknowledgments}
I would like to thank the organizers of the workshop
on Early Physics with heavy--Ions at LHC in Bari for their kind
invitation to this stimulating and pleasant event. 
This work was supported by  
the German Research Foundation under grant GA 1480/2-1.
\end{theacknowledgments}



\begin{thebibliography}{9}

\bibitem{Heinz}
  U.~W.~Heinz, M.~Jacob,
  [nucl-th/0002042].


\bibitem{GaGo}
  M.~Gazdzicki, M.~I.~Gorenstein,
  Acta Phys.\ Polon.\  {\bf B30}, 2705 (1999).
  [hep-ph/9803462].

\bibitem{evidence}
  C.~Alt {\it et al.}  [NA49 Collaboration],
  Phys.\ Rev.\  C {\bf 77}, 024903 (2008)
  [arXiv:0710.0118 [nucl-ex]].

\bibitem{review}
  M.~Gazdzicki, M.~Gorenstein, P.~Seyboth,
  Acta Phys.\ Polon.\  {\bf B42}, 307 (2011)
  [arXiv:1006.1765 [hep-ph]].

\bibitem{proposal}
N.~Antoniou {\it et al.}  [NA61/SHINE Collaboration],
CERN-SPSC-2006-034.

\bibitem{kumar}
  L.~Kumar [ for the STAR Collaboration ],
  [arXiv:1106.6071 [nucl-ex]], \\
  B.~Mohanty [ STAR Collaboration ],
  [arXiv:1106.5902 [nucl-ex]].

\bibitem{schukraft}
  J.~Schukraft {\it et al.} [  for the ALICE Collaboration ],
  [arXiv:1106.5620 [hep-ex]], \\
  A.~Toia {\it et al.} [  for the ALICE Collaboration ],
  [arXiv:1107.1973 [nucl-ex]].

\bibitem{GoGaBu}
  M.~I.~Gorenstein, M.~Gazdzicki and K.~A.~Bugaev,
  Phys.\ Lett.\  B {\bf 567}, 175 (2003)
  [arXiv:hep-ph/0303041].

\bibitem{anar}
  A.~Rustamov, \url{https://indico.cern.ch/conferenceDisplay.py?confId=144745}


\end{thebibliography}
\end{document}